\documentclass[letterpaper]{article} 
\usepackage[submission]{aaai2026}  
\usepackage{times}  
\usepackage{helvet}  
\usepackage{courier}  
\usepackage{amsmath}
\usepackage{amssymb}
\usepackage[hyphens]{url}  
\usepackage{graphicx} 
\usepackage{xcolor}
\usepackage{natbib}  
\usepackage{caption} 
\DeclareMathOperator*{\argmax}{\arg\max}

\usepackage{subcaption}
\usepackage{algorithm}
\usepackage{algorithmic}
\usepackage{newfloat}
\usepackage{listings}
\DeclareCaptionStyle{ruled}{labelfont=normalfont,labelsep=colon,strut=off} 
\floatstyle{ruled}
\newfloat{listing}{tb}{lst}{}
\floatname{listing}{Listing}

\newcommand{\ktau}{\text{Kendall-}\ensuremath{\tau}}

\definecolor{darkred}{rgb}{0.75,0,0}

\title{Personalized Recommendations via Active Utility-based Pairwise Sampling}

\author {
    Bahar Boroomand\textsuperscript{\rm 1},     \,
    James R. Wright\textsuperscript{\rm 2}
}
\affiliations {
    University of Alberta / Amii\\
    Edmonton, Canada\\
    \textsuperscript{\rm 1}bborooma@ualberta.ca,  \, \textsuperscript{\rm 2}james.wright@ualberta.ca
}

\usepackage{bibentry}

\begin{document}

\maketitle

\begin{abstract}
Recommender systems play a critical role in enhancing user experience by providing personalized suggestions based on user preferences. Traditional approaches often rely on explicit numerical ratings or assume access to fully ranked lists of items. However, ratings frequently fail to capture true preferences due to users' behavioral biases and subjective interpretations of rating scales, while eliciting full rankings is demanding and impractical. To overcome these limitations, we propose a generalized utility-based framework that learns preferences from simple and intuitive pairwise comparisons. Our approach is model-agnostic and designed to optimize for arbitrary, task-specific utility functions, allowing the system's objective to be explicitly aligned with the definition of a high-quality outcome in any given application.

A central contribution of our work is a novel utility-based active sampling strategy for preference elicitation. This method selects queries that are expected to provide the greatest improvement to the utility of the final recommended outcome. We ground our preference model in the probabilistic Plackett-Luce framework for pairwise data. To demonstrate the versatility of our approach, we present two distinct experiments: first, an implementation using matrix factorization for a classic movie recommendation task, and second, an implementation using a neural network for a complex candidate selection scenario in university admissions. Experimental results demonstrate that our framework provides a more accurate, data-efficient, and user-centric paradigm for personalized ranking.

\end{abstract}


\section{Introduction}\label{section:Introduction}
Recommender systems play a crucial role in improving user experiences across various domains, including e-commerce, streaming services, and education \cite{veras2015literature}.
Traditionally, these systems have relied on minimizing the difference between predicted and observed numerical ratings \cite{koren2009matrix}. However, these approaches inherently assume that numerical ratings accurately reflect user preferences in a calibrated manner, which is rarely the case in real-world scenarios \cite{mitliagkas2011user}. Miscalibration in user ratings is a well-documented issue in recommendation systems \cite{wang2018your}. Users often rate items inconsistently due to personal rating habits, cognitive biases, or contextual influences.

Rather than predicting numerical scores, the core goal of recommendation is to infer a ranking over items that best matches the user’s underlying preferences. To more directly capture user preferences, some approaches have shifted focus from absolute ratings to relative comparisons.  Pairwise comparisons, in which a system asks which of two items a user prefers, can often provide a more robust signal of user preference than noisy numerical scores.

However, collecting a complete ranking of all items from a user is often infeasible. Instead, real-world preference data typically consists of sparse, partial information \cite{lu2011learning}. Given this inherent incompleteness, we cannot query a user's full preference ordering. Consequently, sampling-based inference methods are necessary to reason about the probability distribution over possible rankings. A central challenge is how to actively and efficiently elicit the most informative preferences from the user to quickly refine this distribution and improve recommendation quality.

This paper introduces a flexible and powerful recommendation framework that learns directly from pairwise comparisons.   Our system directly optimizes the utility of the user, as quantified by an arbitrary utility function.   Thus, our framework can be tailored to prioritize different notions of recommendation quality, such as identifying the single highest-ranked item or ensuring coverage of top-tier candidates.  

To learn user preferences efficiently, we introduce a novel utility-based active sampling method. This strategy adaptively queries the user for the most informative item pairs, i.e., those expected to yield the greatest improvement in the final recommendation utility, thereby accelerating learning and reducing the number of queries needed.  
We use the Plackett-Luce model \cite{plackett1975analysis, luce1959individual}  to represent the distribution over possible rankings given a set of pairwise comparisons that guides the active sampling process.
Our framework is model-agnostic, and to demonstrate its flexibility, we showcase its implementation with two distinct models: a matrix factorization model for a media recommendation task, and a neural network for a complex candidate selection process, such as those found in hiring or university admissions.  In the second experiment, a neural network learns a multifaceted utility function to recommend a shortlist of promising applications, highlighting the system's adaptability to different problem domains and model architectures. Our results confirm that the utility-based framework, combined with active sampling, provides a more accurate, data-efficient, and user-centric recommendation paradigm.

\section{Related Work}\label{section:Related_Wrok}
This section reviews prior work in two key areas. First, we discuss the evolution of recommender systems from traditional rating-based methods to more sophisticated preference-based approaches. Second, we survey relevant techniques for sampling and utility elicitation of preferences, which are essential for learning from partial information.
\subsection{Preference-Based Approaches}

There is widespread recognition in the literature that numerical ratings often fail to accurately represent user preferences. \citet{mitliagkas2011user} highlight that a rating of 7 out of 10, in the absence of context, can be effectively meaningless. Furthermore, both the rating scale and the ratings themselves are often arbitrary and may vary significantly from one user to another, adding to the inconsistency of such measures \cite{ammar2012efficient}. While some methods attempt to extract reliable signals from non-parametrically miscalibrated ratings \cite{wang2018your} or propose hybrid models combining ratings with comparisons \cite{bledaite2015pairwise}, empirical evidence strongly favors a complete shift away from rating-based systems. Studies show that direct preference judgments are not only faster for users to provide but also approximately $20\%$ more stable over time, underscoring their methodological advantages \cite{carterette2008here, jones2011comparisons}. This has motivated a move toward models that learn directly from comparisons.

Foundational work in this paradigm includes constructing user profiles by defining similarity through modified cosine metrics tailored to preference structures \cite{brun2010towards} or using probabilistic latent class models where preferences within each latent community are governed by a Bradley-Terry model \cite{bradley1952rank, liu2009probabilistic}. A highly influential framework is Bayesian Personalized Ranking (BPR) \cite{rendle2012bpr}, which offers a principled method for learning from implicit feedback by creating pairwise preferences from user interactions and optimizing a generic, Bayesian-derived pairwise loss function. In the related listwise domain, models like ListNet \cite{cao2007learning} learn ranking functions by minimizing the cross-entropy between predicted and ground-truth top-one probabilities, a concept later integrated with matrix factorization by ListRank-MF \cite{10.1145/1864708.1864764}. While these frameworks establish the value of learning from passively collected data, our approach is distinguished by its use of a utility-driven strategy for actively selecting informative comparisons.

Other ranking methods have also been proposed, such as EigenRank \cite{liu2008eigenrank}, which aggregates neighbor preferences via a random walk after identifying them using Kendall’s 
$\tau$ rank correlation \cite{kendall1938new,marden1996analyzing}, but its reliance on online neighbor searches poses scalability challenges. Similarly, specialized matrix factorization techniques have been designed to learn latent user and item-pair representations for improved \emph{precision at $k$}, a common measure of ranking performance \cite{kalloori2016pairwise}, yet they still minimize score discrepancies, inheriting biases from rating-based formulations. Even a recent deep learning architecture that employs autoencoders and blends pairwise loss with rating reconstruction retains a dependency on the same biased ratings \cite{10.1145/3018661.3018720} . In contrast, our framework operates purely on comparisons, leveraging the Plackett-Luce model to capture full ranking distributions and actively querying the user to optimize arbitrary, task-specific utility.

\subsection{Active Elicitation for Preference Learning}

A key challenge in preference learning is actively eliciting information from users. A seminal approach in this area is the decision-theoretic framework of \citet{chajewska2000making}, who proposed modeling a user's utility function as a random variable and iteratively selecting queries with the highest myopic value of information (VOI). This foundational idea was extended by \citet{boutilier2002pomdp}, who framed the elicitation task as a Partially Observable Markov Decision Process (POMDP) to enable non-myopic, sequential query planning, representing a more theoretically robust but computationally complex alternative. A different paradigm for active selection is based on racing algorithms, which aim to efficiently identify the best items with high statistical confidence. \citet{busa2013top} generalized racing to work with noisy pairwise preferences and later, \citet{busa2014preference} tailored these active sampling strategies specifically for learning Mallows models, a distance-based probability distribution over rankings \cite{mallows1957non}. By assuming that user preferences are generated from an underlying Mallows distribution, they exploit the model’s statistical properties to devise querying algorithms. 

Learning from the partial preference data gathered through such elicitation requires sophisticated inference techniques. A significant body of work has focused on learning specific families of statistical ranking models. \citet{lu2011learning} developed algorithms for learning the Mallows model from arbitrary pairwise comparisons, using novel sampling methods for inference within a Monte Carlo EM algorithm. The Bayesian Mallows model was later applied by \citet{liu2019diverse} to learn from implicit feedback, demonstrating that a principled probabilistic approach can improve recommendation diversity and provide valuable uncertainty estimates. While these methods provide rigorous solutions for learning the Mallows model from static data (either explicit pairs or implicit clicks), our framework is grounded in the related Plackett-Luce model.  This model defines ranking probabilities as a direct function of individual item scores, a structure that is well-suited for the efficient, iterative updates needed to evaluate potential queries in our active loop. The distance-based formulation of the Mallows model, by contrast, makes repeated calculations computationally prohibitive for an interactive system. While the Plackett-Luce model provides a principled statistical foundation, the core contribution of our work is not the model itself, but rather the general, utility-driven, and active learning architecture that is built upon it.

\section{Pairwise Comparison Model}\label{section:Pairwise_Comparison_Model}
Most recommender systems operate by learning a scoring function, $s_{ui} = f(u, i; \Theta)$, which estimates the preference of a user $u$ for an item $i$, based on a set of model parameters $\Theta$. In traditional systems, this score is trained to predict an explicit numerical rating, and the model's parameters are optimized to minimize the error (e.g., MSE) between the predicted scores and the observed ratings. However, the ultimate goal is to generate an accurate ranking of items, not to precisely predict ratings. Small differences in predicted ratings may not significantly impact the MSE but can result in an incorrect ordering of items, which is critical for user satisfaction. For instance, for a user $u$, item $i$ is rated $r_{ui} = 4.32$ and item $j$ is rated $r_{uj} = 4.24$, but the model predicts scores $s_{ui} = 4.27$ and $s_{uj} = 4.30$. In this scenario, the error between the actual rating and the predicted one is relatively small, but the order of items is inverted. This highlights that minimizing rating errors does not guarantee optimal ranking performance. 

An alternative and more direct approach is to model preferences using pairwise comparisons. Instead of relying on numerical ratings, this method uses data of the form $i \succ_u j$, indicating that user $u$ prefers item $i$ over item $j$. This reframes the learning problem from rating prediction to modeling the probability of observed choices, thereby avoiding the biases inherent in subjective rating scales.

\subsection{Pairwise Loss Function}

A powerful framework for modeling preferences from ranked data is the Plackett-Luce model \cite{plackett1975analysis, luce1959individual}. The model's central assumption is that each item $i$ from a universal item set $\mathcal{I}$ possesses a positive, latent utility score.

The probability that a user prefers one item over another is proportional to their respective scores. Given a ranking of $K$ items, $\pi = (\pi_1, \pi_2, \ldots, \pi_K)$, the probability of the ranking $\pi$ given the vector of scores $\theta$ is:
\begin{align}
    \label{equation_plackettLuce}
    P(\pi \mid {\theta}) = \prod_{k=1}^{K} \frac{\theta_{\pi_k}}{\sum_{j=k}^{K} \theta_{\pi_j}}.
\end{align}
In many applications, preferences are personalized. We therefore define $\theta_{ui}$ as the user-specific utility score that user $u$ assigns to item $i$. The learning objective is to infer these latent scores based on observed preference data.

For a dataset composed of triplets $(u, i, j)$, where user $u$ prefers item $i$ to item $j$, the probability of a single observed comparison under the Plackett-Luce model is:
\[ P(i \succ_u j \mid \Theta) = \frac{\theta_{ui}}{\theta_{ui} + \theta_{uj}}\]
Here, the latent score $\theta_{ui}$ is generated by the underlying model. For numerical stability, it is common to work with log-scores, such that $s_{ui} = \ln \theta_{ui} = f(u, i; \Theta)$. The function $f$ can be any model, from a simple dot product to a complex neural network.

The optimization objective is to maximize the log-likelihood of the observed comparisons given the model's parameters $\Theta$. The loss function is:
\begin{align}
    \mathcal{L}(\Theta; \{(u,i,j)\}) &= \sum_{(u,i,j)} \ln P(i \succ_u j \mid \Theta) \notag\\
    &= \sum_{(u,i,j)} \ln \frac{\exp(s_{ui})}{\exp(s_{ui}) + \exp(s_{uj})}.
    \label{equation_loss}
\end{align}

To optimize the parameters $\Theta$ using gradient-based methods, one can compute the gradient of the loss with respect to the model's output scores, $s_{ui}$ and $s_{uj}$. For a single triplet $(u,i,j)$:

\begin{align}
    \frac{\partial \mathcal{L}}{\partial s_{ui}} = 1 - \sigma(s_{ui} - s_{uj}) = \frac{\exp(s_{uj})}{\exp(s_{ui}) + \exp(s_{uj})},
\end{align}

\begin{align}
    \frac{\partial \mathcal{L}}{\partial s_{uj}} = \sigma(s_{ui} - s_{uj}) - 1=  \frac{- \exp(s_{ui})}{\exp(s_{ui}) + \exp(s_{uj})},
\end{align}
where $\sigma(\cdot)$ is the sigmoid function. A detailed analysis of how our pairwise approach fundamentally outperforms traditional rating-based prediction is provided in the Appendix.

\section{Utility-Based Recommendation}\label{section:Utility-Based_Recommendation}
Traditional rating-based models aggregate ratings and recommend items that have the highest average ratings to the user.  However, the object of interest is a user's ranking over items, not the items' ratings. 
%
In our approach, the outcome of the model's training is not a direct prediction of item ratings but rather a posterior probability distribution over the user's preference rankings.
This enables direct optimization of the criteria for what constitutes a ``good'' recommendation.
Typically, a user will not select every item that has been recommended to them; in fact, frequently they will select only one.  Thus, the average ``rating'' of the recommended items is of no importance; a recommendation is ``good'' if, on average, it contains at least some highly-preferred items.  For instance, if a user’s preferences are ambiguous, it might be better to recommend a few items that are highly preferred by different groups to which a user might belong, rather than a set of items that are most likely to be preferred by the average user.

We directly encode the notion of a ``good'' recommendation via \emph{utility functions} $U$, which numerically evaluate the quality of a \emph{menu} $m$ of recommended items.  We say that a menu $m$ is a better recommendation than $m'$ for a user with a preference ordering $\pi$ whenever $U(m,\pi) > U(m',\pi)$.
Since a user's preference order will, in general, be unknown, we will typically operate on the expected utility $\bar{U}$, defined as:
\begin{equation}
    \bar{U}(m \mid \theta) = \mathbb{E}_{\pi} \left[ U(m \, ,\pi) \right] = \sum_{\pi \in \Pi} P(\pi | \theta) \times U(m \, ,\pi),
\end{equation}
where $P(\pi \mid \theta)$ is the probability of a specific ranking $\pi$ over items given model parameters $\theta$. As the model trains, it adjusts its ranking probabilities $P(\pi \mid \theta)$ accordingly, leading to a progressive refinement in recommendation quality as the model becomes more confident about the user's preferences through continued pairwise interactions.

We choose recommendations $f_u(\theta)$ that have high expected utility for the user.  Formally:
\begin{equation}
\label{equation_f}
    f_u(\theta) = \argmax_{{m} \in A^k} \,  \bar{U}(m \, ,\, \theta),
\end{equation}
where $A^k$ is the set of every $k$ possible recommendations.

\subsection{Utility Function Instantiations}

A fundamental advantage of our framework is its flexibility in accommodating various utility functions depending on the specific applications.  In this subsection we describe two examples of utility functions for different application domains.

\paragraph{Goods and Media.}
In media-related applications such as streaming services or e-commerce platforms, the primary objective is to recommend a small set of highly relevant items.\footnote{Even when a large list is presented to the user, it is well-established that items higher on the list are much more likely to be interacted with, leading to a small effective limit on the number of items that can be shown \cite{joachims2007evaluating}.}
Typically, a user will choose only a single item from the recommendation; thus, it is  more important to improve the expected quality of the single most-preferred item in the recommendation than it is to improve the expected average quality of over all the recommended items.  These criteria are encoded by the utility function
\begin{equation}
\label{media utility}
    U(m, \pi) = \max_{j \in m} \, \mathrm{rank}(j; \pi).
\end{equation}
Here $j$ refers to the items in the menu $m$, and $\mathrm{rank}(j; \pi)$  represents the position of $j$ within the permutation $\pi$. The utility $U(m, \pi)$  measures the highest rank achieved by any item $j$  in the recommendation set $m$  within a specific ranking permutation $\pi$.

\paragraph{Admissions.}
In some applications, the important question is how many of the overall top-ranked items were included in the recommendation \cite{liu2009probabilistic}, rather than the rank of the single highest-ranked item included.  For example, when choosing a cohort of students to admit, a university might prefer to admit as many of the top $k$ applicants as possible, rather than to optimize the rank of the top applicant that they admit.  In such a scenario, the best menu is one which includes the largest fraction of the top-$k$ items in the preference ordering, so a more appropriate utility function would be
\begin{equation}
    U(m, \pi) = \sum_{i=1}^{k} \mathbb{I}[\pi_i \in {m}],
    \label{eq:admissions}
\end{equation}
where $\mathbb{I}[\pi_i \in m]$  is an indicator function that is $1$ when item $\pi_i$ (the $i$-th highest item in the ranking) is included in the set $m$, and $0$ otherwise.

\section{Utility-Based Active Sampling}\label{section:Utility-Based_Active_Sampling}
In practical applications, complete user rankings are rarely available; instead, data typically consists of partial preference information \cite{lu2011learning}. Consequently, sampling-based inference is required to reason about the underlying preference distributions from this incomplete information.
In our model, active sampling is used to further enhance the learning process by selectively querying the most effective item pairs. Unlike passive models that rely on user interaction data as it accumulates, active sampling strategically selects pairs of items for which user feedback would be most valuable in improving the model's understanding of preferences and then explicitly queries the users to rank the pairs.
For pairwise comparisons, active sampling is crucial because not all item pairs provide the same level of information. By focusing on the most informative comparisons, the model accelerates learning and improves recommendation accuracy, particularly when data is sparse or in cold start scenarios.

Our method takes an end-to-end approach to active sampling by integrating it directly into the optimization framework, rather than relying on proxy methods such as entropy or vector norms \cite{rashid2002getting,rubens2015active}.  These methods, designed to quantify uncertainty, do not directly quantify the metric of interest.  In particular, not all uncertainty is equally valuable to resolve; information that does not change the system's choice of recommendation contributes little.

Our approach chooses a query $g(\theta)$ that optimizes \emph{utility gain} derived from the pair comparison, as follows.
\begin{align}
    \label{equation_g}
    g(\theta) = \arg\max_{(i,j)\in A^2} \mathbb{E}_\theta \ [U(f(\theta|q), \theta|q) - U(f(\theta), \theta|q)]
\end{align}
Here $\theta$ is the the set of beliefs (model parameters or factor matrices),
$A^2$ is the set of possible pairwise queries, and $q$ is the possible outcomes for the query $(i,j)$ result,
and $\theta|q$ are the model parameters that would result from retraining with result $q$ added to the training set.  The first term of the expectation represents the expected quality of the recommendation that we would choose after seeing $q$, taking expectation with respect to the beliefs updated for $q$.  The second term represents the quality of the recommendation that we would choose \emph{before} seeing $q$, but also taking expectation with respect to the updated beliefs.  This difference is evaluated for both possible outcomes $q$, with the outer expectation weighting the two according to their probability of occurence according to the pre-query beliefs.

The pair that maximizes this expected utility gain is selected for querying the user, and the model is updated according to the results of the query. This strategy ensures that the system actively samples the pairs that are most likely to improve the utility of recommendations, rather than treating all reductions of uncertainty equally.

It can be computationally expensive to compute the expectations in \eqref{equation_g}, particularly in large-scale applications, as there are $O(n!)$ possible permutations of $n$ items.  Rather than exhaustively consider all possible orderings, we approximate these expectations using Monte Carlo sampling.

\subsection{Monte Carlo approximation}
\label{Monte Carlo subsection}
Rather than summing over all possible rankings, we can efficiently sample a set of rankings  $\pi^{(1)}, \pi^{(2)}, \ldots, \pi^{(R)}$  from the distribution  $P(\pi | \theta)$ using the Plackett-Luce model \cite{xia2019learning}.
For each sampled ranking  $\pi^{(r)}$, we calculate the utility contribution $U(m,\pi^{(r)})$.
Once the utility values are computed for all  $R$  sampled rankings, we approximate the overall utility by averaging these values:
\begin{align*}
    \bar{U}(m, \theta) \approx \frac{1}{R} \sum_{r=1}^{R} \left( U(m, \pi^{(r)}) \right)
\end{align*}

This approximation allows us to estimate the utility function without computing over the entire set of $n!$  possible rankings, making the approach computationally feasible for large datasets.
Since each sampled ranking contributes to our utility based only upon the top $k$ items in the ranking, we further reduce computational requirements by only sampling the first $k$ items in each preference ordering, treating all other items identically as having rank $n$.  This is without loss of generality for the utility defined in \eqref{eq:admissions}, since only the top $k$ items of any ranking influence a menu's utility.
See Algorithm~\ref{alg:monte_carlo_utility} for a detailed definition.

\begin{algorithm}[t]
\caption{Monte Carlo  for Utility Approximation}
\label{alg:monte_carlo_utility}
\begin{algorithmic}[1]
\STATE \textbf{Input:} $m$, $\theta$, $R$, $k$, $n$
\STATE \textbf{Output:} Approximation of $\bar{U}(m, \theta)$
\STATE \textbf{Initialize:} $U_{total} \leftarrow 0$

\FOR{$r = 1$ \TO $R$}
    \STATE Initialize score set $S \leftarrow \{s_1, s_2, \ldots, s_n\}$
    \STATE $\triangleright$ \textit{Sample $\pi^{(r)}$ from the Plackett-Luce model:}
    \FOR{$i = 1$ \TO $k$}
        \STATE Sample item $\pi^{(r)}_i$ from $S$ with probability:
        \[
        P(\pi^{(r)}_i | S) = \frac{s_{\pi^{(r)}_i}}{\sum_{j \in S} s_j}
        \]
        \STATE Remove $\pi^{(r)}_i$ from $S$.
    \ENDFOR
    
    \STATE $\triangleright$ \textit{Compute utility: $U(m, \pi^{(r)})$ }
    \STATE $utility \leftarrow 0$
    \FOR{$i = 1$ \TO $k$}
        \IF{$\pi^{(r)}_i \in m$}
            \STATE $utility \leftarrow utility + 1$
        \ENDIF
    \ENDFOR

    \STATE $U_{total} \leftarrow U_{total} + utility$
    
\ENDFOR

\STATE $\triangleright$ \textit{Normalize the utility:}
\[
\bar{U}(m, \theta) \leftarrow \frac{U_{total}}{R}
\]

\RETURN $\bar{U}(m, \theta)$
\end{algorithmic}
\end{algorithm}

\section{Experiments}\label{section:Experiments}
To validate our utility-based framework, we conduct two distinct experiments. The first demonstrates the effectiveness of our utility-based active sampling strategy in a classic media recommendation scenario using the Matrix Factorization (MF) models. The second showcases the framework's flexibility by applying it to a candidate selection task using Neural Networks (NN).

\subsection{Experiment 1: Media Recommendation with MF}
This experiment evaluates the core contribution of our active sampling method. The goal is to show that querying a user for a small number of intelligently selected preference pairs can lead to significant improvements in recommendation quality.

\subsubsection{Data and Preprocessing}
We use the MovieLens 100k dataset, a benchmark containing 100,000 ratings from approximately 1,000 users across 1,700 movies \cite{harper2015movielens}. To simulate a real-world scenario where labelled data is scarce, 
the model is pretrained with $5$ randomly chosen comparisons for each user. This reflects a cold-start problem where little user interaction data is available. After the pertaining phase, the active learning strategy draws pairs to query the user, and the responses to queries are simulated based on ground-truth preferences derived from the dataset. 

For the final evaluation, a dedicated test set of items is held out. To accurately assess the model's ability to generalize to novel content, these test items are excluded from the sampling pool.

\subsubsection{Model Implementation: Matrix Factorization}
For this experiment, the general scoring function $f(u, i;\Theta)$ is instantiated as a Matrix Factorization model, a widely used approach in recommender systems due to its ability to capture user-item interactions in a low-dimensional latent space. The preference score that user $u$ gives to item $i$ is modelled by the dot product of their latent vectors:
\begin{align*}
    s_{ui} = {u}_u^\top {v}_i
\end{align*}
Here, ${u}_u, {v}_i \in \mathbb{R}^K$ are $K$-dimensional latent feature vectors for the user and item, respectively. We set $K=100$. The model parameters $\Theta=\{U,V\}$ are optimized using the pairwise loss function defined in Equation \ref{equation_loss}.

\subsubsection{Experimental Protocol}
We compare three approaches to demonstrate the value of our active sampling strategy:
\begin{enumerate}
    \item Baseline (No Sampling): A standard pairwise comparison model trained only on the initial training set.
    \item Random Sampling: The baseline model is augmented by querying the user for their preference on randomly selected item pairs from the sampling pool.
    \item Utility-Based Active Sampling: (Our proposed method) It actively selects item pairs to query by optimizing for the expected improvement in recommendation utility, using the Goods and Media utility function (Equation \ref{media utility}). To maintain computational tractability, we use a Monte Carlo approximation to estimate the expected utility gain.
    
\end{enumerate}
For each query, the newly acquired preference pair is added to the training set, and the model is retrained.

\subsubsection{Results} 
\begin{figure}[h!]
\centering
\includegraphics[width=1.0\linewidth]{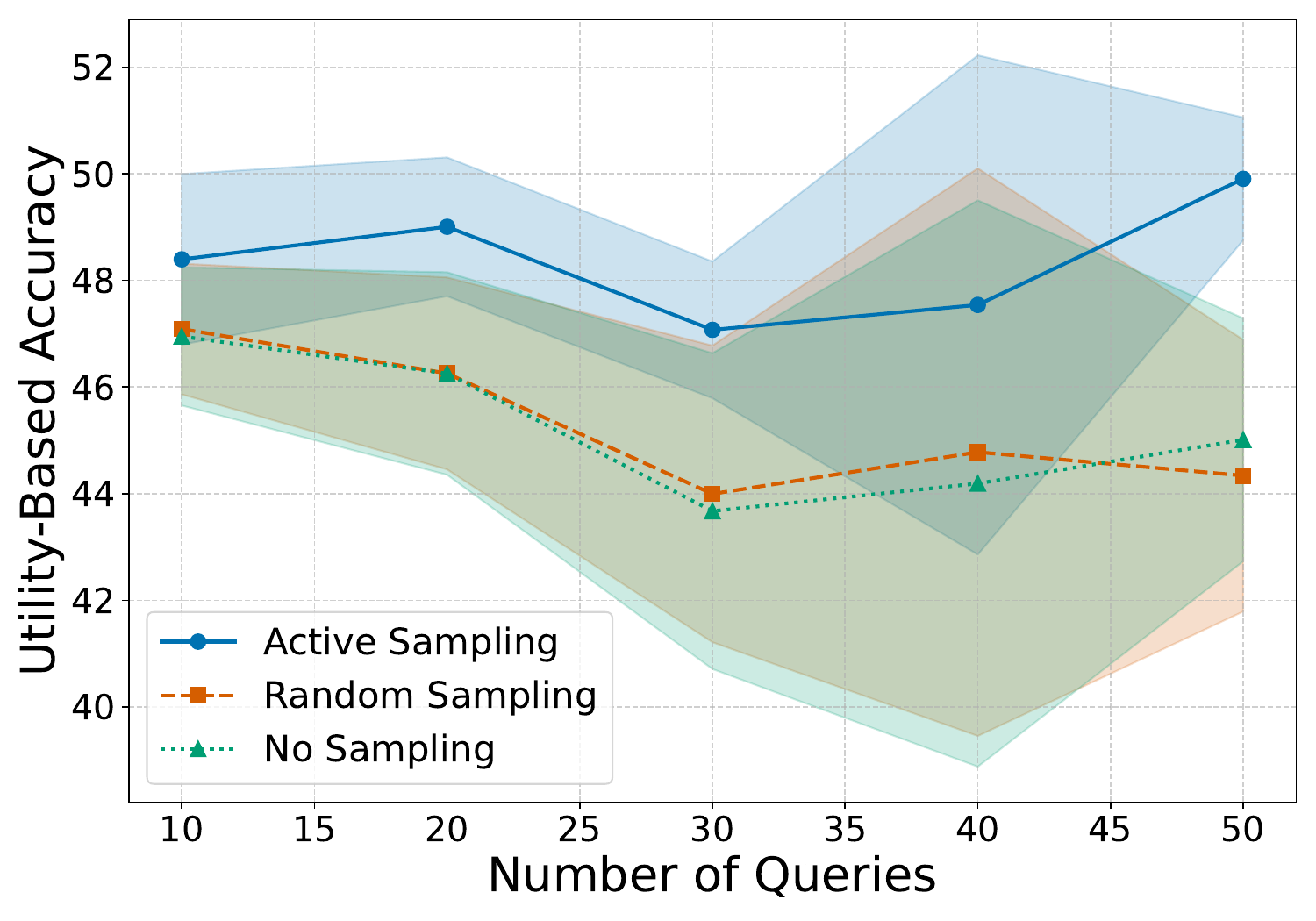}
\caption{Comparing the standard pairwise model with no sampling as baseline, the pairwise model with random Query selection, and the pairwise model with utility-based active sample average over multiple runs}
\label{fig:Active_sampling}
\end{figure}

To assess model performance, we evaluate the quality of top-$k$ recommendations.
After training and potentially querying, each model is called upon to produce a menu of recommendations for each user. The only items available for recommendation are those in the test set (i.e., those that did not appear in any comparison in either the sampling set or the training set).  This is intended to correspond to the media application: users can only rank items that they have previously selected (e.g., movies that they have already watched).  But recommendations should only include items that the users have \emph{not} previously selected.
A predicted menu is rated more highly if its max-rank item is higher among the user's ground truth ranking over all test items. For ease of interpretation, we divide this number by  $k$  to obtain percentages, which are represented on the y-axis in Figure \ref{fig:Active_sampling}.

Figure \ref{fig:Active_sampling} shows the result of the comparison between the three models.
Note that although the no-sampling baseline model (green line) does not query,
its performance is different for different numbers of queries.  This is because we re-sample and re-train each model in each scenario.

In this experiment, we start our queries from $10$ users per epoch (which is approximately $1\%$ of the users in the dataset) and increase the number of queries incrementally. Our utility-based active sampling consistently outperforms both random sampling and the baseline (no sampling) across all numbers of sampled items, as shown by the higher curve and confidence interval for the blue line. This gap becomes more pronounced as the number of sampled items increases, even for very small quantities of additional data.
With only 50 additional queries \emph{total} (an average of $0.05$ queries per user), we see a significant improvement in recommendation performance, demonstrating a dramatic improvement in data efficiency.

\subsection{Experiment 2: Candidate Selection with NN}
This experiment demonstrates the framework's flexibility and effectiveness in a high-stakes decision-making context, such as hiring or university admissions. We shift from the media domain to an admissions task, replace the previous model with a neural embedding architecture, and utilize active querying to maximize performance with limited feedback.

\subsubsection{Dataset}
We utilize a dataset for graduate admissions based on the work of \citet{acharya2019comparison}, available at \citet{acharya_graduate_admissions}, which is derived from real-world admissions statistics. The dataset contains records for 400 applicants, each described by eight features: GRE Score, TOEFL Score, University Rating, Statement of Purpose Score, Letter of Recommendation Score, Undergraduate GPA, and Research Experience. The final feature, ``Chance of Admission'', provides a continuous score from 0 to 1, which we use to establish a ground-truth ranking of all candidates.  The objective is to efficiently identify the top-$k$ candidates from this ground-truth ranking, minimizing the number of preference queries posed to the decision-maker.

\subsubsection{Model Implementation}
Recent work has demonstrated the effectiveness of neural network models for recommendation and preference learning \cite{he2017neural,song2018neural, he2018outer}.
Our approach adopts the Neural Network architecture (NeuMF) model proposed by \citet{he2017neural}, which is well-suited for capturing complex, non-linear relationships in our oracle-candidate interactions. 
While the original work explored pointwise optimization methods, our model uses the pairwise preference loss function defined in Equation \ref{equation_loss}. This architecture serves as the base model, on top of which we implement our utility-based active sampling strategy.
\\
The NeuMF model combines the outputs of two distinct pathways:
\begin{enumerate}
    \item Generalized Matrix Factorization (GMF): This path uses an element-wise product of the preference and candidate embedding vectors, allowing it to learn linear interaction patterns.

    \item Multi-Layer Perceptron (MLP): This path concatenates the preference and candidate embeddings and processes them through a series of non-linear hidden layers, enabling it to model more complex and higher-order relationships.
\end{enumerate}
The outputs from the GMF and MLP paths are concatenated and passed through a final output layer to produce the comparisons between candidates. To quantify model uncertainty for our active learning strategies, we implement a technique inspired by epistemic neural networks
\cite{dwaracherla2024efficient}. Specifically, we introduce a latent random vector $z$, drawn from a standard normal distribution, which is projected and added to the core preference embedding.

\subsubsection{Experimental Protocol}
The experiment compares the efficiency of our Utility-Based Active Sampling strategy against three baseline approaches. The protocol is as follows:
\begin{enumerate}
    \item Utility-Based Active Sampling: (Our proposed method) It actively selects item pairs to query by optimizing for the expected improvement in recommendation utility, using the Admissions utility function (Equation \ref{eq:admissions}).

    \item Entropy (Uncertainty Sampling): A classic active learning baseline which selects the pair that model is most uncertain about the relative ranking. This is measured by the predictive entropy on the binary outcome of the comparison. 

    \item  Random Sampling: As a simple baseline, it randomly select a pair from the set of all possible pairs. This strategy is non-active and does not use any information from the model to guide its selection

    \item Cluster-Based Sampling: This baseline serves to test whether a structured but non-adaptive sampling strategy can match active learning. It pre-generates a sequence of queries using K-Means clustering approach, ensuring a diverse set of comparisons.
    
\end{enumerate}
The experiment simulates an active learning process over $100$ iterations. At the beginning of each iteration, the model's current performance is evaluated and querying strategy selects a pair of candidates $(i,j)$ for comparison. The ground-truth is used to simulate an oracle's preference (e.g., $i \succ_u j$), and this new pair is added to the training set. Finally, the NeuMF model is fine-tuned on this newly augmented dataset, completing the loop and preparing the model for the next iteration.

\subsubsection{Results}
To evaluate the effectiveness of our proposed active learning strategy, we employ two standard ranking metrics, Precision@10 and Normalized Discounted Cumulative Gain (NDCG)@10. Precision@10 measures the fraction of relevant items within the top-$10$ recommended candidates. It is a measure of set accuracy, indicating the number of correct candidates identified within the recommendation slate. NDCG@10 is a metric of ranking quality that considers the position of candidates in the recommended list. It assigns higher scores for placing more relevant candidates at higher ranks, making it sensitive to the overall ordering of the results.

\begin{figure}[h]

    \begin{subfigure}{1.0\linewidth}  
        
        \includegraphics[width=\linewidth]{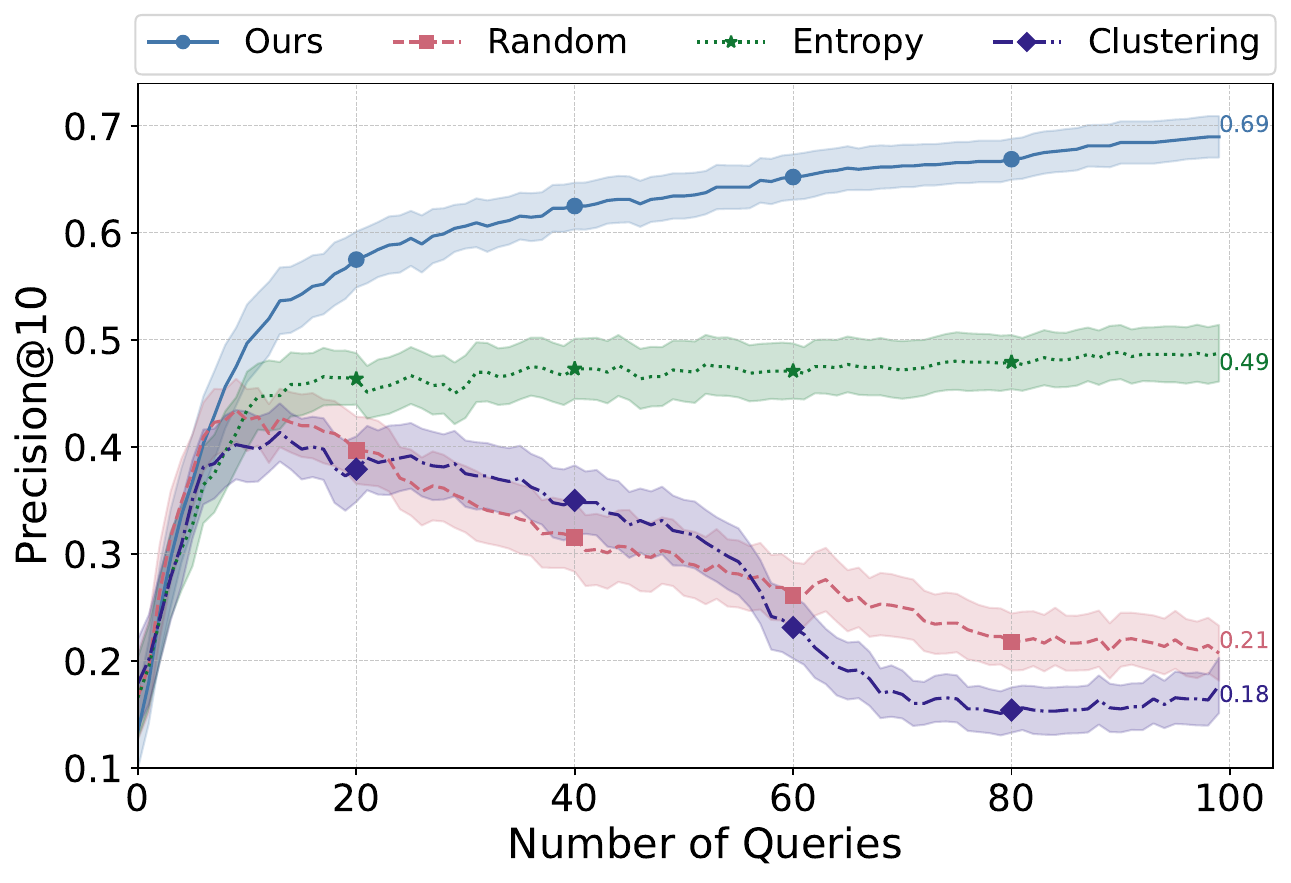}
    \end{subfigure}
    ~\\
   \begin{subfigure}{1.0\linewidth}
        \includegraphics[width=\linewidth]{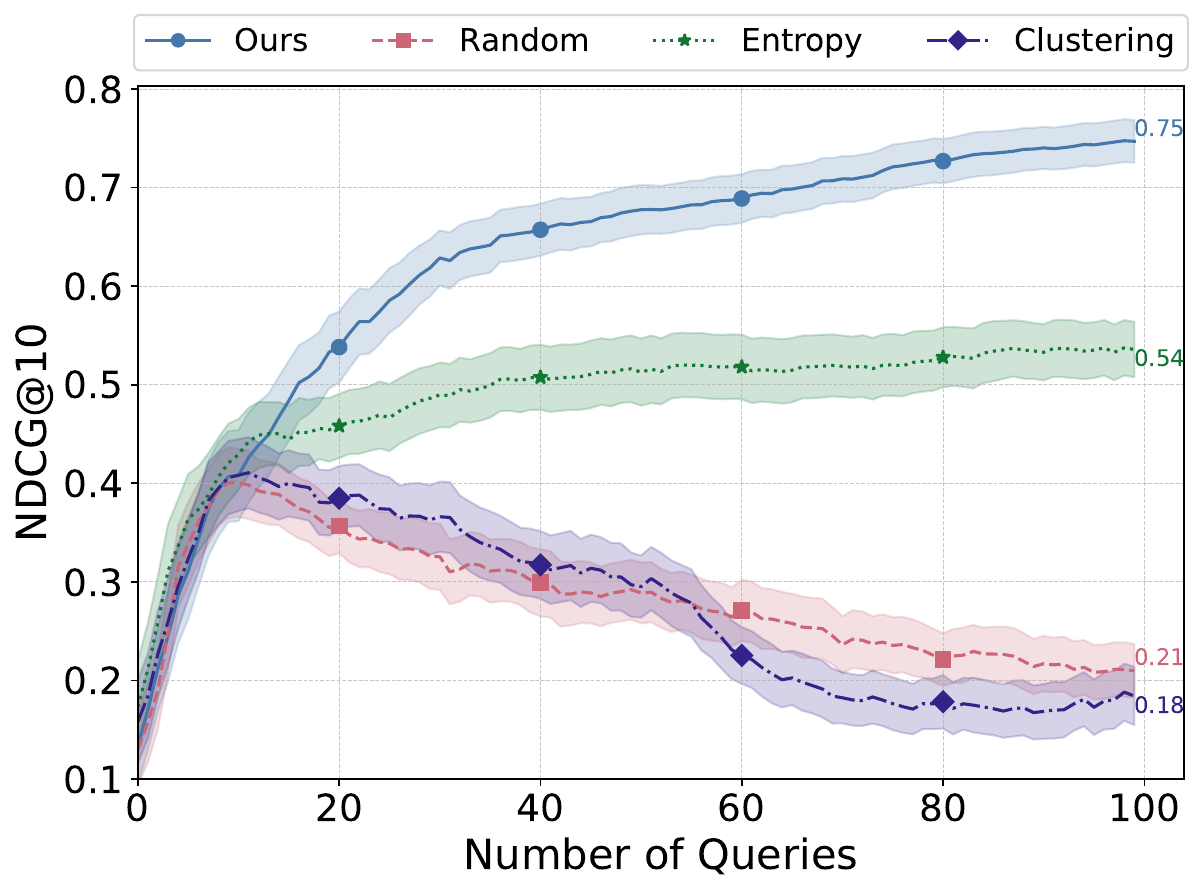}

\end{subfigure}
\caption{Comparing our pairwise utility-based active sampling model with Entropy-based sampling, Random sampling and Clustering as baseline, average over 100 runs}
\label{fig:Admissions}
\end{figure}

Figure \ref{fig:Admissions} shows Precision@10 and NDCG@10, both as functions of the cumulative number of pairwise-comparison queries issued to the recommender. In every case, our approach rises more rapidly and finishes markedly higher than the three baselines. These curves demonstrate that actively selecting the most informative comparisons not only accelerates the early-stage learning of user preferences but also delivers substantially superior ranking accuracy once the annotation budget is exhausted, validating the effectiveness of our proposed framework.

\section{Conclusion}\label{section:Conclusion}
In this work, we introduced a flexible and powerful recommendation framework that learns user preferences directly from pairwise comparisons. By moving beyond the limitations of traditional rating models, our approach directly models the underlying goal of recommendation, which is to produce an optimal ranking of items for a user. The core of our framework is the integration of explicit utility functions and a novel utility-based active sampling strategy. This active learning component adaptively queries users for the most informative comparisons, maximizing the utility gain and accelerating the learning process. 

Our empirical results demonstrate the effectiveness of this framework across two distinct domains. In a media recommendation task using a Matrix Factorization model, our method significantly outperformed the random sampling baseline, showing remarkable data efficiency even with a very small number of queries. In a candidate selection scenario using a Neural Network, our approach again surpassed multiple baselines, including entropy-based sampling, in both precision and NDCG. These results confirm that our approach of  directly optimizing for recommendation utility and actively seeking the most valuable user feedback enables   more accurate, efficient, and user-centric recommendations.

\bibliography{aaai2026.bib}

\clearpage
\appendix 
\section{Appendix}\label{section:appendix}
\label{Appendix}
To establish the foundational advantage of modelling preferences directly, we conducted a semi-synthetic experiment comparing our preference-based model against a traditional rating-based Matrix Factorization model.
\subsubsection{Datasets}
We based our comparisons on MovieLens, a commonly used benchmark recommendation dataset containing 1 million ratings from 6000 users on 4000 movies \cite{harper2015movielens}. Each record includes the user ID, movie ID, and user rating.
We divide the dataset into training and test sets with different train sizes \{0.1\%, 1\%, 10\%,20\%,40\%,60\%,80\%\}, enabling analysis of the impact of the training set size on both approaches. 

The core insight that this work exploits is that ratings are not a ground truth observation.  Thus, for our evaluation, we pre-process the ratings data to construct a semi-synthetic ground truth set of preference orders, as follows. A low-rank matrix factorization model is trained on the entire dataset, yielding user and item factor estimates ${u}_u$ and ${v}_i$. These are used to construct ``shadow ratings'' via 
$\hat{r}_{ui} = {u}_u^\top {v}_i$; these will be the ratings that are reported to the rating-based matrix factorization algorithm.  The shadow ratings are then used as the parameters of a Plackett-Luce model, from which each user's ground truth preference ordering is sampled.  The probabilities of the model are adjusted by Laplace smoothing (i.e., addition of a small constant $\alpha$ to each item's pre-normalization probability) to enforce a minimum probability for each item, as below:
\begin{equation}
P'(i) = \frac{\theta_i + \alpha}{\sum_{j \in S} (\theta_j + \alpha)}
\end{equation}

\subsubsection{Experimental Protocol}
We have two approaches to
demonstrate the value of our pairwise comparison approach:
\begin{itemize}
    \item Normal Matrix Factorization (Rating-Based):
    This method minimizes the squared difference between predicted and observed shadow ratings, with L2 regularization, as in \citet{koren2009matrix}.
\[
\min_{{u}, {v}} \sum_{(u,i)\in \text{train}} (r_{ui} - {u}_u^\top {v}_i)^2 + \lambda \|{u}\|^2 + \lambda \|{v}\|^2.
\]
We train using stochastic gradient descent (SGD), adjusting user factors ${u}_u$ and item factors ${v}_i$ each epoch.

\item Pairwise Comparison Model (Preference-Based): Our proposed method optimizes item rankings by learning from pairwise outcomes. For each preference $i \succ j$, we use the log-likelihood loss defined in equation~\eqref{equation_loss}, based on observed preference comparisons $(u, i, j)$.

\end{itemize}
In both methods, the user and item factor matrices, $U$ and $V$, are randomly initialized from a normal distribution. The priors for the latent factors are defined as independent distributions:
\begin{align}
        P({U}) =  \prod_{i=1}^{N_U}\prod_{k=1}^K \mathcal{N}(u_{ik}; 0, 1),
    \\
 P({V}) =  \prod_{j=1}^{N_V} \prod_{k=1}^K\mathcal{N}(v_{jk}; 0, 1).  
\end{align}

Here $K$ is the number of features and $N_U$ is the number of users, and $N_V$ is the number of items. Throughout the experiments, we set the number of features to $100$.

\begin{figure}[t]
    \includegraphics[width=1.0\linewidth]{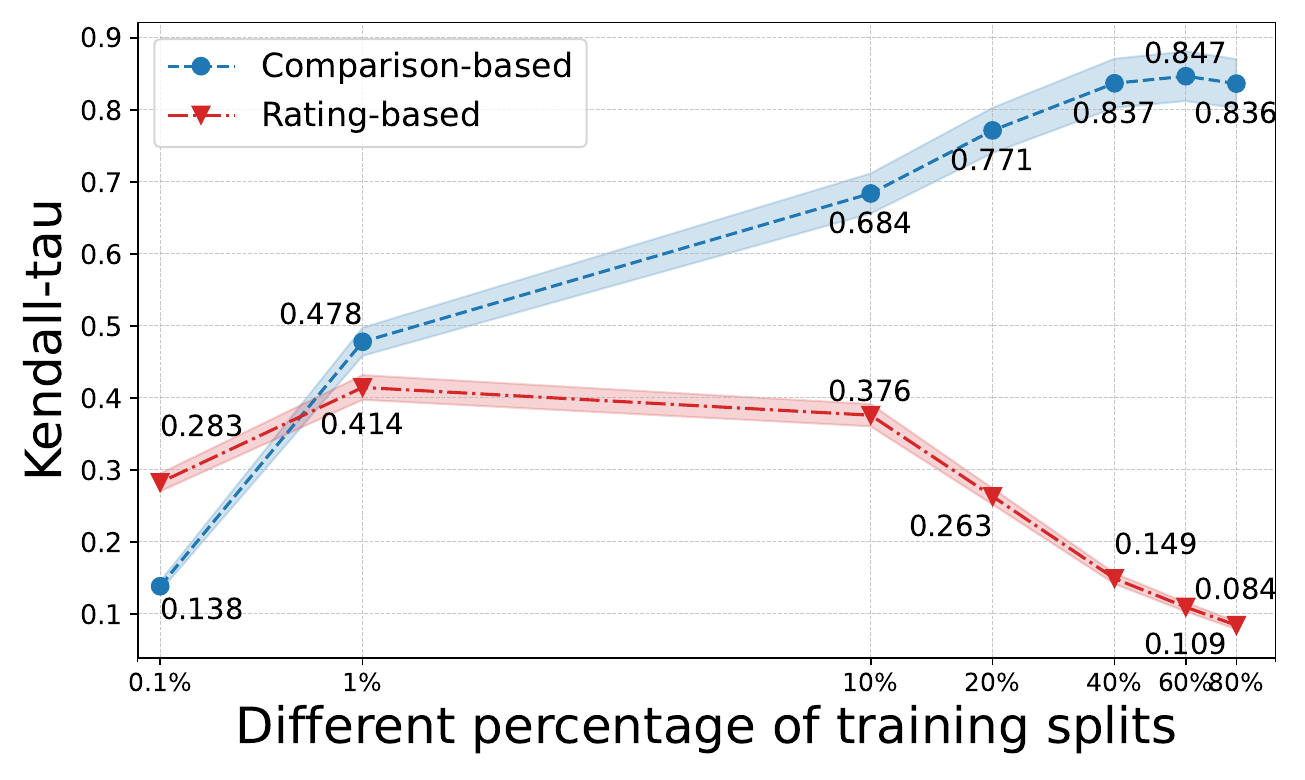}
\caption{\ktau{} results across different training set fractions for comparison-based (blue) and rating-based (red) models, averaged over 100 runs.}
\label{fig:KendallTauResults}
\end{figure}

\begin{figure*}[h!]
\begin{subfigure}{1\linewidth}
    \includegraphics[width = \linewidth]{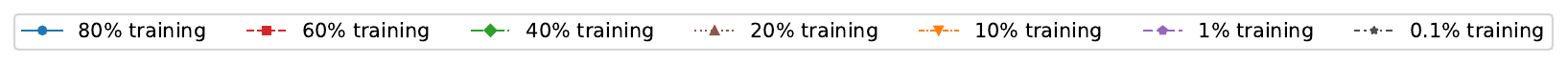}
\end{subfigure}\\
\begin{subfigure}{0.48\linewidth}  
        \includegraphics[width=\linewidth]{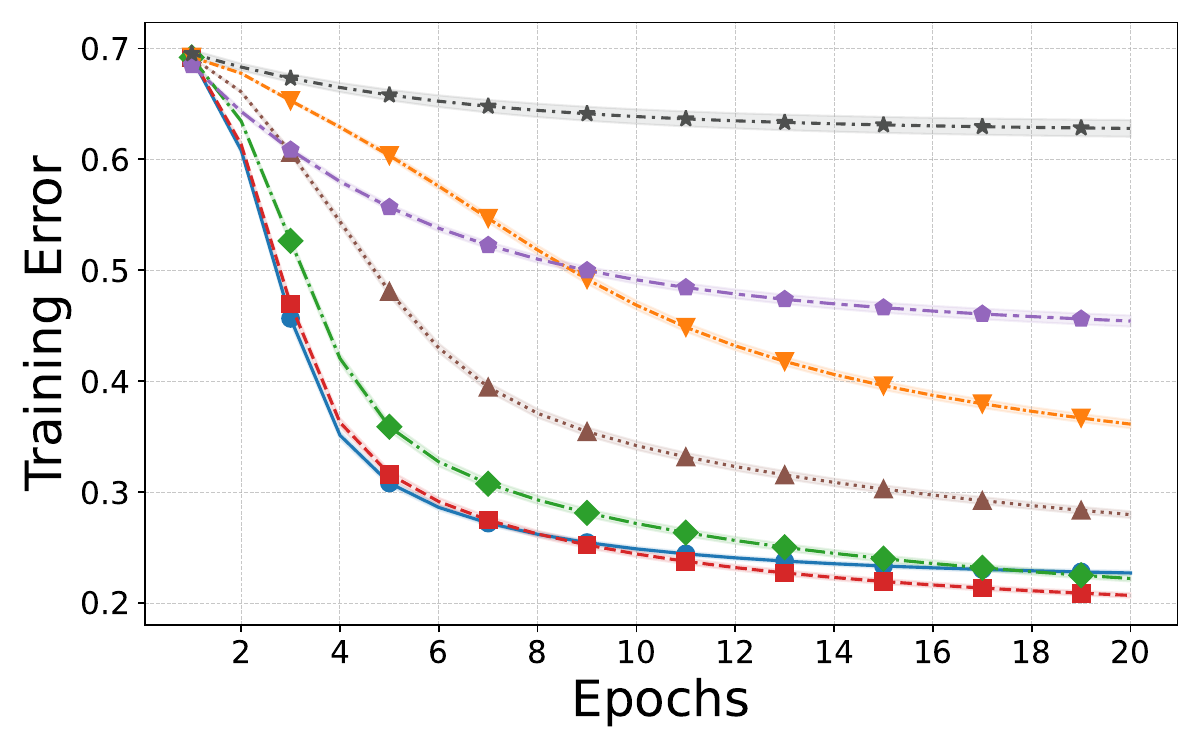}
        \caption{Training error for comparison-based algorithm across different training splits}
    \end{subfigure}
    \hfill
    \begin{subfigure}{0.48\linewidth}
    
        \includegraphics[width=\linewidth]{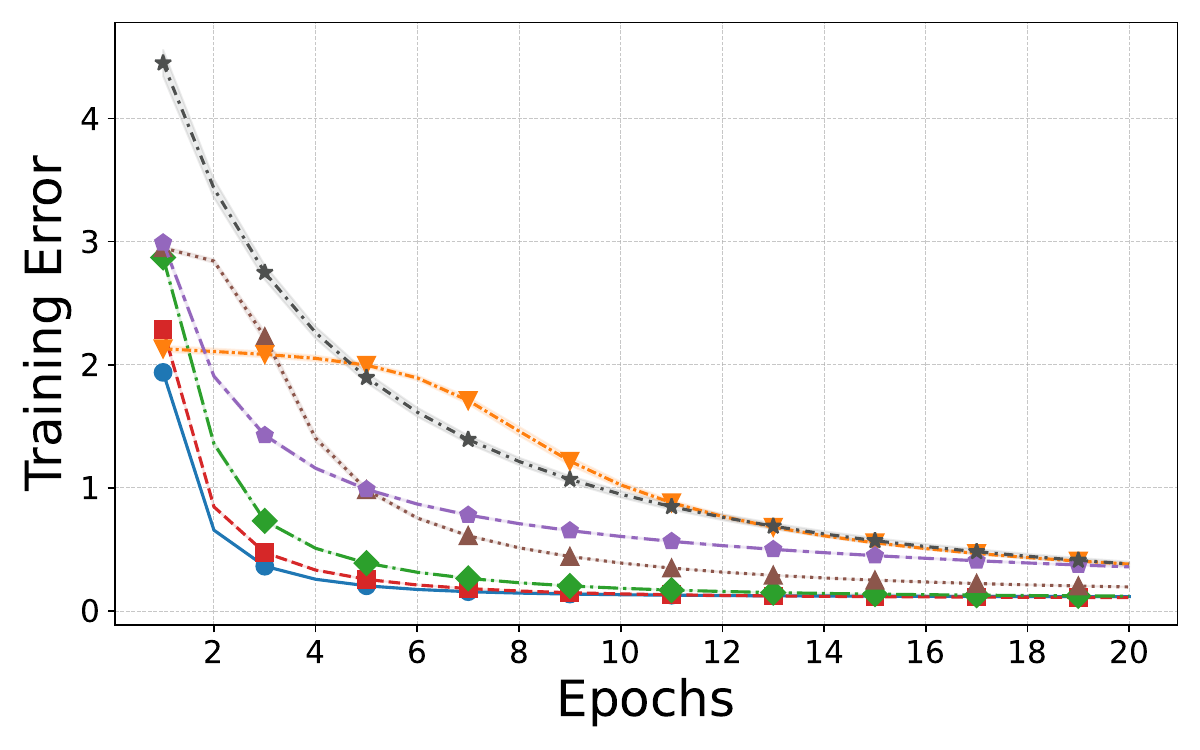}
        \caption{Mean Squared Error for rating-based algorithm across different training splits}
    \end{subfigure}
    
    \caption{Error of the training process for the comparison-based and the rating-based, averaged over 100 runs}
    \label{fig:Training_MSE}
\end{figure*}

\subsubsection{Results}
For these experiments, we measure performance using \ktau{}, a rank correlation coefficient that quantifies the ordinal association between two ranked lists \cite{kendall1938new,marden1996analyzing}.
\[
\tau = \frac{P - Q}{\sqrt{(P + Q + T)(Q + Q + U)}}.
\]
Here, $P$ is the number of concordant pairs, $Q$ the number of discordant pairs, $T$ the number of ties only in the first list, and $U$ the number of ties only in the second one. Given two sets of rankings, \ktau{} measures how frequently two items are ordered identically in both lists. The coefficient ranges from $-1$ (perfect negative correlation) to $+1$ (perfect positive correlation), with $0$ indicating no significant correlation. A higher \ktau{} value thus corresponds to a stronger agreement in item ordering.
This metric is particularly well-suited for recommender systems because it directly measures how closely two rankings agree, which aligns closely with how users experience recommendations. Rather than focusing on the precise accuracy of numerical ratings, \ktau{} evaluates whether items users truly prefer are ranked higher than those they like less.

In practice, user satisfaction in a recommender system depends heavily on whether the system places the most relevant (or preferred) items at or near the top of the user's recommended list \cite{cremonesi2010performance}.
Users are indifferent to small variations in predicted ratings, for example, 4.2 instead of 4.9; however, they are sensitive to the order in which items are recommended, particularly when a less preferred item is ranked ahead of a more preferred one \cite{cremonesi2010performance}. Consequently, rating differences are not a reliable metric for evaluating recommendation quality, whereas the ranking order is a more meaningful indicator of user satisfaction.  Since \ktau{} quantifies the proportion of \emph{pairs} of items ranked correctly, it effectively captures the user-centric notion of prioritization. The higher the \ktau{} coefficient, the more the system's ordering aligns with the user's true preference order, and thus the greater the potential user satisfaction.

Figure \ref{fig:KendallTauResults} shows that our comparison-based model consistently achieves much higher Kendall-$\tau$ than the rating-based model, except at the tiniest fractions of the training set.

Notably, the ranking-based model's performance degrades as the training sample grows larger than $10\%$ of the dataset size, unlike the comparison-based model, whose performance improves with larger quantities of training data.  At the same time, the comparison-based model is robust to reductions in the training set size, producing strong \ktau{} even at lower fractions of the dataset.
This robustness is particularly valuable for real-world recommender systems, where acquiring large, high-quality labelled datasets may be costly or impractical.

Figure~\ref{fig:Training_MSE} shows that both models steadily reduce their training error with larger training sets.  Thus, the rating-based model appears to learn a misaligned objective, as improvements in its training objective do not translate into meaningful improvements in ranking quality in Figure~\ref{fig:KendallTauResults}.

\end{document}